\documentclass{acm_proc_article-sp}
\begin{document}
\title{A Vertical Channel Model of Molecular Communication based on Alcohol Molecules}

%
%
%
%
%

\numberofauthors{4} 
%
\author{
%
%
\alignauthor
PengFei Lu\\
   	   \affaddr{School of Computer Science}\\
       \affaddr{Shaanxi Normal University}\\
       \affaddr{Xi'an, Shaanxi, China}\\
       \email{jacinto@snnu.edu.cn}
\alignauthor
Yang You\\
        \affaddr{School of Computer Science}\\
       \affaddr{Shaanxi Normal University}\\
      \affaddr{Xi'an, Shaanxi, China}\\
       \email{youyang@snnu.edu.cn}
\and  
\alignauthor
Bo Liu\\
        \affaddr{School of Systems Information Science}\\
       \affaddr{Future University Hakodate}\\
       \affaddr{Hokkaido, Japan}\\
       \email{boliujsj@gmail.com}
\alignauthor
ZhenQiang Wu\\
       \affaddr{School of Computer Science}\\
       \affaddr{Shaanxi Normal University}\\
       \affaddr{Xi'an, Shaanxi, China}\\
       \email{zqiangwu@snnu.edu.cn}
}



\maketitle
\begin{abstract}
The study of Molecular Communication(MC) is more and more  prevalence, and channel model of MC plays an important role in the MC System. Since different propagation environment and modulation techniques produce different channel model, most of the research about MC are in horizontal direction,but in nature the communications between nano machines are in short range and some of the information transportation are in the vertical direction, such as \emph{transpiration} of plants, \emph{biological pump} in ocean, and \emph{blood transportation} from heart to brain. Therefore, this paper we propose a vertical channel model which nano-machines communicate with each other in the vertical direction based on pure diffusion. We first propose a vertical molecular communication model, we mainly considered the gravity as the factor, though the channel model is also affected by other main factors, such as the flow of the medium, the distance between the transmitter and the receiver, the delay or sensitivity of the transmitter and the receiver. Secondly, we set up a test-bed for this vertical channel model, in order to verify the difference between the theory result and the experiment data. At last, we use the data we get from the experiment and the non-linear least squares method to get the parameters to make our channel model more accurate.
\end{abstract}
%
\section{Introduction}
Nowadays, modern telecommunication system conveys most information with electrical or electromagnetic signals. However, still there are many applications which are not convenient or appropriate for these technologies. For example, the use of wireless communication inside networks of tunnels, pipelines, or salt water environment, can be very inefficien \cite{2014-Farsad-p-}. As another example, with the dimensions of the transmitter and receiver become smaller and smaller, electromagnetic communication is extremely challenging because of constrains such as the ratio of the antenna size to the wavelength of the electromagnetic signal \cite{2008-Akyildiz-p2260-2279}.
\\
Inspired by nature, one possible solution to these problems is to use chemical signals as carriers of information, which is called \textit{molecular communication}(MC) \cite{2013-Nakano-p-}. In molecular communication, a transmitter releases small particles such as molecules or lipid vesicles into an aqueous or gaseous medium, where the particles propagate until they arrive at a receiver. The receiver then detects and decodes the information encoded in these particles.Moreover, molecular communication signals are biocompatible, and require very little energy to generate and propagate. These properties makes chemical signals ideal for many applications, such as biomedical application, environment application, where the use of electromagnetic signals are not possible or not desirable.
\\	
	Although MC is present in nature and is used by micro-organisms, it was only recently that engineering a MC system has been proposed as means of communication at the microscale \cite{1999-Freitas-p-, 2006-Hiyama-p162-162}. Macroscale MC was not even considered until Dr.Nariman Farsad developed the first macroscale test-bed for MC \cite{2013-Farsad-p-a}.	
\\
In \cite{2013-Farsad-p-a}, Dr.Nariman Farsad developed the first horizontal, modular, and programmable platform for MC in macro-scale, which makes a big leap in the development of Molecular Communication. Nariman et al. elaborate on the horizontal flow assisted propagation of molecular communication in a long range, but discusses little about the short range of MC based on pure diffusion. In biological systems, the communications between nano machines are in short range in most situation, also particles(information molecules) are precious resources, thus releasing large numbers of particles at once is impractical, we need to consider limitations on the transmission amount of the molecules. Also for the reason that in the vertical direction, the receiver can receive the peak value \cite{2011-Garralda-p443-448} of the information molecules. In nature there are many examples of information transportation in the vertical direction, such as \emph{transpiration} \cite{1958-Wit-p-} of plants, \emph{biological pump} \cite{2001-Ducklow-p50-58} in ocean, biological nitrogen removal using a vertically moving biofilm system \cite{2004-Rodgers-p313-319}, and blood transportation from heart to brain. Therefore, we need to consider both pure diffusion and information transmission in the vertical direction.
\\
In this paper, we propose a channel model of the isopropyl alcohol molecules' propagation in the vertical direction. Also we refer Nariman Farsad's model to set up our new platform to test the channel model we proposed. The test bed we developed only measures the peak of the transmission, which is especially useful in biological system.
\\
This paper is organised as follows. In Section \ref{section_MC_via_diffusion}, we describe the Molecular Communication via Diffusion. In Section \ref{section_vertical_model}, we describe the theory model and set up a test-bed of molecular communication model. In Section \ref{section_fitting_model}, we first analyse the theory knowledge of MC via Diffusion, and then through experiments, we showed the difference of experiment result with the theory result, at last we use the non-linear least squares curve fitting method to get the parameters of the coefficients. Last in Section \ref{section_conclusion}, we draw a conclusion of our work.
\section{Molecular Communication via Diffusion}\label{section_MC_via_diffusion}
We model a communication system composed of a pair of devices, each called a nanonetworking-enabled node(NeN,ie.,nano node or nano robot)\cite{2014-Yilmaz-p929-932}. In Molecular Communication via Diffusion(MCvD), the NeNs communicate with each other through the propagation of certain molecules via diffusion\cite{2010-Pierobon-p602-611,2014-Pierobon-p2085-2095}. The Molecular Communication via Diffusion is depicted as Fig. \ref{fig:MC_sketch}.
\begin{figure}[H]
		\centering		
		\epsfig{file=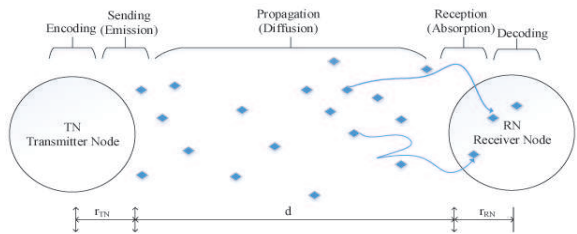,width=3in,height=1.2in}
		\caption{MCvD system model in a 3-D environment\cite{2014-Yilmaz-p929-932}}
		\label{fig:MCvD model}
\end{figure}
The MCvD system has five main processes: encoding, sending, propagation, reception, and decoding\cite{2012-Kuran-p65-73,2015-Tepekule-p1311-1314}. During the encoding process before modulating the symbols during the sending process, may need some channel coding techniques. After emitting the messenger molecules, the propagation process plays an important part in transmitting the messenger molecules to the receiver. When some messengers molecules arrive at the receiver(ie., hit the receiver), they will form chemical bonds with the receptors on the surface of the receiver. And after the molecular messengers hit the receiver they are removed from the channel and the properties of these received molecules(e.g., concentration, type)constitute the received signal. Demodulation function takes place during the absorption process and it is followed by decoding if a channel coding is employed.
\\
In a MCvD system, communication pairs are assumed to be synchronized. Information is modulated on some of the physical properties of messenger molecules, it can be the number, type, or any other property of the messenger molecules. The information messengers' movement is modulated as either Brownian motion or diffusion process.
The motion is governed by the combined forces due to thermal energy of the medium. If we consider the diffusion process of particles starting from its transmitter, then the concentration at distance $d$ and time $t$ is given as 
	\begin{equation}\label{formular_pulse_n_dimission}
		f(d,t)=\frac{M}{(4{\pi}Dt)^{n/2}}\exp\left[\frac{-d^2}{4Dt}\right]\cite{2014-Farsad-p-} \quad t > 0 \mbox{ and } d >0\enspace.
	\end{equation}		
where $n$ and $D$ are the dimension of the environment and the diffusion coefficient\cite{2001-Redner-p-}, and $M$ stands for the amount of molecules. $D$ depends on temperature of the environment, viscosity of the fluid, and Stokes' radius of the molecules\cite{1984-Tyrell-p-}.
\section{A vertical Model of Molecular Communication}\label{section_vertical_model}
\subsection{Theory Model of the vertical Model}\label{section_new_model}
Since different propagation scheme and modulation technique lead to different channel model. Therefore, we propose a new channel model, which looks like the following Fig. \ref{fig:MC_sketch}, when in a vertical situation, the receiver can receive the peak value \cite{2011-Garralda-p443-448} of the information molecules, which is very effective for communication, and in biological, the information molecules spread for every direction, and vertical diffusion is one of the common phenomenon, nutrition is transmitted in bloods of human or animals and pheromones are transmitted in plants, both are also in this direction sometimes, therefore, it's necessary to research it.
\begin{figure}[H]
	\centering
	\epsfig{file=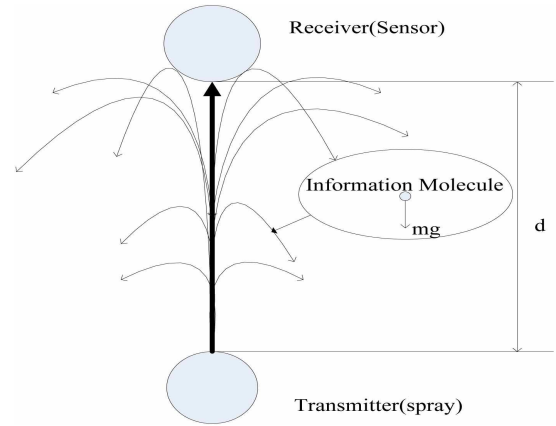,width=3in,height=2.2in}
	\caption{The sketch model of Molecular Communication}
	\label{fig:MC_sketch}
\end{figure}
For this situation, we consider the amount of the molecules at different time steps, and when the information molecules are transmitted in the vertical direction, the channel is basically affected by the transmitter and receiver's delay, the flow of the medium, also we need to consider the gravity of the mass molecules which we denote $mg$ and we all known $mg$ is positive.
\\
In the uniform acceleration motion,if we know the initial velocity, the instant time , the we get the velocity of the entity at time $t$ from the following equation 
\begin{equation}
v(t)=v_0+at \quad t \geq 0 \enspace .
\end{equation}
where $v(t)$ is the velocity of the entity at time $t$, $v_0$ is the initial velocity, $a$ is the acceleration, and $a$ is either negative or positive.
\\
 Similarly, we assume that the decrease of molecules is proportion to gravity as time gone, also due to the flow may affect the gravity, we denote the final result of gravity as $e$,  the amount of the molecules which denotes as $f(t)$ is at different time steps can be regard as the velocity of the entity at time $t$, then we get the following equation
\begin{equation}\label{formular_before_new_model}
	f(t)=C(t)-et \quad t > 0 \enspace .
\end{equation}
where $C(t)$ looks like the initial velocity in the uniform acceleration, but it decreases with time .
\\
 Due to the first part $C(t)$, we should introduce several parameters to adjust the formula of (\ref{formular_pulse_n_dimission}) in 1-D\cite{2014-Farsad-p2392-2401}, $n$ equals 1, then we will get the following formula
\begin{equation}\label{formular_new_model}
f(t)=\frac{a}{\sqrt{t}}\exp\left[{\frac{-bd^2}{t}}\right]-et \quad t > 0 \enspace .
\end{equation}
where $d$ is the distance between sender and receiver, and $t$ is the time, for parameter $a$ corresponds to this part $\frac{M}{\sqrt{4\pi{D}}}$ of (\ref{formular_pulse_n_dimission}), which is related to the diffusion coefficient and the delay of transmitter and receiver, and $b$  corresponds to the part $\frac{1}{4D}$ of (\ref{formular_pulse_n_dimission}), which is affected by the diffusion coefficient, for the parameter $e$ is related to the diffusion coefficient and the gravity of molecules, also we assume that the variation of gravity is a constant. For the value of the coefficients, we will discuss in Section. \ref{Fitting model}.
\subsection{A test-bed of Molecular Communication}\label{section_Real_Model}
\begin{figure}[H]
		\centering
		\epsfig{file=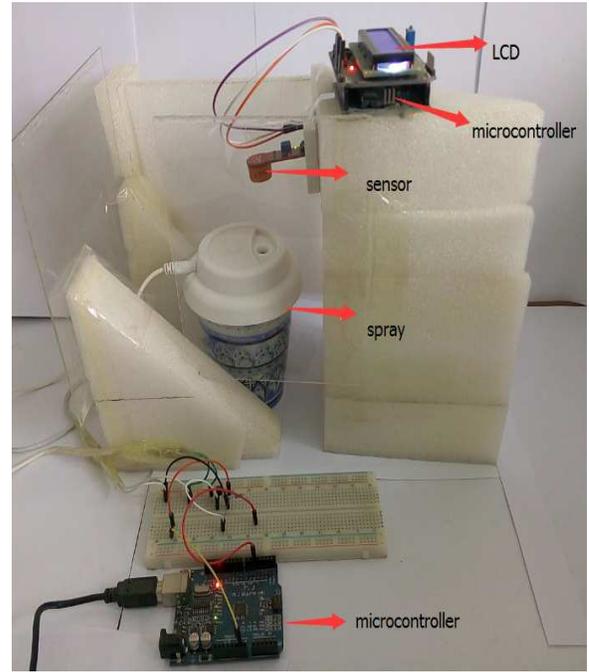,height=3.5in,width=3in}
		\caption{A test-bed of Molecular Communication}			
		\label{fig_real_model}
\end{figure}

\begin{figure}[H]
		\centering
		\epsfig{file=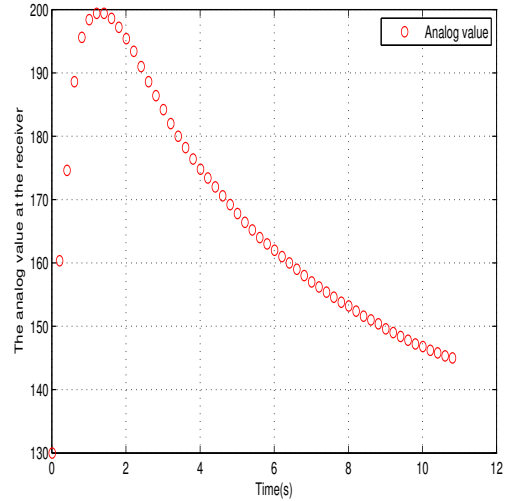,height=2.85in,width=3in}
		\caption{The analogy value at the receiver vs Time}
		\label{fig_single_exp}
\end{figure}

In this section, we introduce a test-bed of molecular communication which is depicted as Fig. \ref{fig_real_model}\cite{2015-You-p-}. Our test-bed refer to Dr.Nariman Farsad's work\cite{2013-Farsad-p-a,2015-Kim-p-}.
The transmitter is composed of a spray and a microcontroller. When an input command is given to the microcontroller, the information is converted into a binary stream, which can be transmitted through different modulation schemes by precisely controlling the spray. The chemical signals is used as the information. When the spray releases these molecules, they propagate through the medium(i.e.,air) until reach the receiver. The receiver consists of an alcohol sensor and a microcontroller that reads the sensor data which displays on the LCD screen. Since alcohol is used as carrier of information, MQ-3 semiconducting metal oxide gas sensor is used for detection at the receiver. The microcontroller at the receiver side reads the sensor data using an analogy to digital converter. The data can then be analyzed and sent to a computer serial port. It was shown that short text messages could be transmitted across a room using this setup through concentration shift keying\cite{2011-Kuran-p1-5}. In this work, we analyse the system response of the platform of this test-bed.
To perform some measurements, we separate the transmitter and the receiver by 10 cm. 
Fig. \ref{fig_single_exp} shows the analogy value of alcohol molecules at the receiver side. We wait between each trial until the initial analogy reading of the sensor drops to about a fixed number. Although it is extremely difficult to find the exact cause of deviations between trials, some likely causes are: the spray, which is not precise to spray the same amount of alcohol for each trial; the flow can be turbulent; the sensor can be noisy; and other environment factors such as random flows in the room.	
\section{Experiment Model and Fitting Model}\label{section_fitting_model}
	In this section, we present the difference of the theoretical models and the experimental data from our platform, also according to the experiment data and non-linear least squares method, we fitted a more accurate model for our test-bed.
\subsection{Models Versus Experimental Results}\label{section_model_vs_exp}
\begin{figure}[H]
		\centering
		\epsfig{file=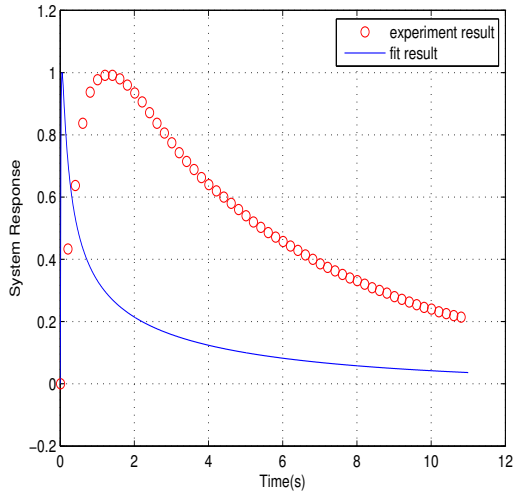,height=2.8in,width=3in}
		\caption{Comparison of the experimental data and theoretical models}
		\label{fig_exp_theory_simulation}
	\end{figure}

In this section we compare the previously published theoretical models to the experimental results obtained using the table-top platform. To demonstrate this, we set the distance of  the transmitter and the receiver is 10 cm. Table \ref{table_system_param} summarizes all the experiment parameters we will use in our test-bed.
\begin{table}[h]\scriptsize
\caption{The Experiment Parameters}
\centering
\begin{tabular}{ll}
\hline\noalign{\smallskip}
Parameters & Values \\
\noalign{\smallskip}
\hline
Distance between a transmitter and a receiver & 10cm \\
Spraying during for each bit & 100ms \\
Diffusion coefficient of isopropyl alcohol & 0.0993cm${^2}$/s\cite{1968-Lugg-p1072-1077} \\
Temperature(room temperature) & 25$^{\circ}$C=298K \\
\hline
\end{tabular}
\label{table_system_param}
\end{table}	
\\If these parameters are used in the theoretical formula, the system response can be calculated. Because the number of particles released by the transmitter is not known, we assume $M=1$ and then normalize the plots by dividing them by their respective maximum values. Similarly the system responses obtained from experiment result is normalized with its maximum. By normalizing the plots, we compare only the shape of the theoretical results with the shape of the experimental results. For our experimental system response, we average the response of 6 different experimental trials to produce a single plot. As Fig. \ref{fig_exp_theory_simulation}, shows the theory simulation versus the experiment simulation at different time steps. 
	According to the shape of the two curves, we find that the theoretical model reaches its peak early than the experiment model. Then we need to consider some delay in the experiment model. The sensitivity of transmitter and receiver and the flow of the medium may lead to the delay. we may reduce the delay when we adjust the formula of the theoretical by adding or adjust some parameters, we may get a curve as the experimental curve. For the details we will discuss in the Section. \ref{Fitting model}.
\subsection{Fitting model}\label{Fitting model}
\begin{figure}[H]
		\epsfig{file=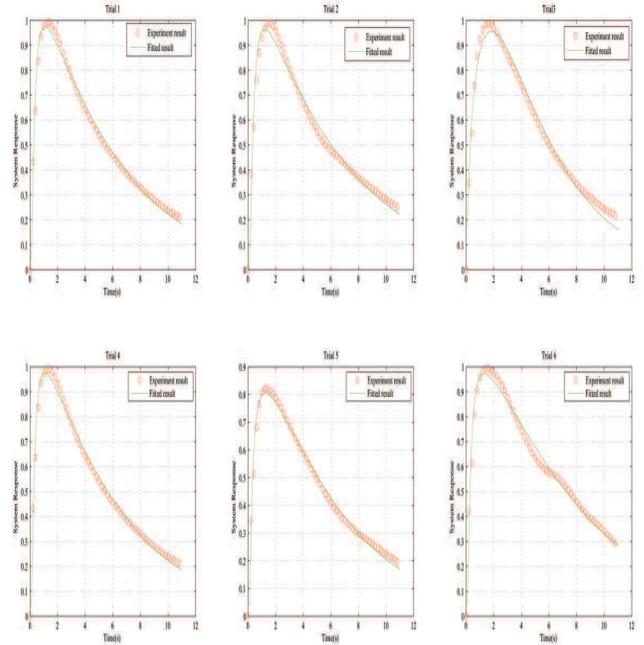,height=3.5in,width=3.35in}
		\caption{System response and the fitted model for set of 6 trials}
		\label{fig_six_trials}
	\end{figure}

\begin{figure}[H]
		\centering
		\epsfig{file=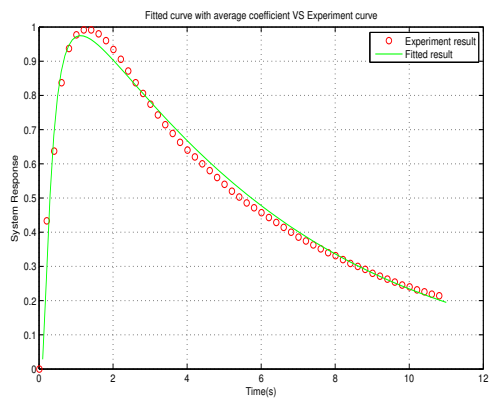,height=2.2in,width=3in}
		\caption{The fitting line vs the experiment line}
		\label{fig_fit_exp}
	\end{figure}

	For the reason that only the first few seconds of the pulse response is typically used in practice for information transmission\cite{2013-Farsad-p-a} in the molecular communication. Also, in order to get a better and more accurate result, we use the first 11 seconds of sensor measurements to fit the curve of the proposed model.
	\\
	Through the experiment data we got from our test-bed, we use non-linear least square curve fitting  method \cite{1963-Marquardt-p431-441} to estimate the coefficients of $a$, $b$, $e$. In Fig. \ref{fig_six_trials} shows the 6 trials of the experiments and the fitting models from the data we get. And the red dot is the data from the test-bed, and the green line is the fitting models. 
\noindent	After six trials of experiments, we choose coefficients of $a$, $b$, $e$ as the average of each trials. Then we get coefficients of $a$, $b$, $e$ are 1.8788,60.4567,0.0301 respectively. Then we get the following formula from(\ref{formular_new_model}). 
	\begin{equation}\label{formular_mend_end}
		f(t)=\frac{1.8788}{\sqrt{t}}\exp\left[\frac{-60.4567d^2}{t}\right]-0.0301t \quad t > 0\enspace .
	\end{equation}
\\
Again we do the experiment with the coefficients we get, and plot the results as the Fig. \ref{fig_fit_exp}.
\section{Conclusions}\label{section_conclusion}
	In this paper, we showed a vertical MC channel model based on the alcohol molecules. From Fig.\ref{fig_exp_theory_simulation}, we know that the experiment results have some difference from the theory results, but through comparing the model (\ref{formular_new_model}) to the experiments results, we can get the coefficient of the new model(\ref{formular_mend_end}) through experiments with non-linear least squares method. As the Fig.\ref{fig_single_exp} show, the variation of the analogy value with time fits the power low after the maximum value.\\
	For future work, we will research the difference of theoretically from the experiment results, how the distance between transmitter and receiver affect on the efficiency of MC, also its application in blood transmission from heart to brain.
\section{Acknowledgments}
   This work was supported by the Fundamental Research Funds for the Central Universities(GK201402038,GK201501008) and the National Natural Science Foundation of China(61173190).

\end{document}